\documentclass[preprint,preprintnumbers,nofootinbib,amsmath,amssymb]{revtex4-1}
\usepackage{amsfonts}    
\usepackage{amssymb}
\usepackage{amsmath}
\usepackage{latexsym}
\usepackage{eepic}
\usepackage{graphicx}
\usepackage[usenames,dvipsnames]{color}
\usepackage{color}

\usepackage[active]{srcltx}
\usepackage{epstopdf}

\newcommand{\al}{\alpha}
\newcommand{\pa}{\partial}

\newcommand{\Om}{\Omega}
\newcommand{\om}{\omega}

\newcommand{\De}{\Delta}

\newcommand{\tha}{\theta}

\newcommand{\rar}{\rightarrow}

\begin{document}

\title{From two-dimensional (super-integrable) quantum dynamics to (super-integrable) three-body
dynamics}

\author{Alexander V Turbiner\\[8pt]
Instituto de Ciencias Nucleares, UNAM, M\'exico DF 04510, Mexico\\[8pt]
turbiner@nucleares.unam.mx\\[8pt]
     Willard Miller, Jr.\\[8pt]
School of Mathematics, University of Minnesota, \\
Minneapolis, Minnesota, U.S.A.\\[8pt]
miller@ima.umn.edu\\
[10pt]
and \\[10pt]
M.A.~Escobar-Ruiz,\\[8pt]
Departamento de F\'isica,
Universidad Aut\'onoma Metropolitana-Iztapalapa, San Rafael Atlixco 186,
M\'exico, CDMX, 09340 M\'exico\\[8pt]
admau@xanum.uam.mx}

\begin{abstract}
It is shown that planar quantum dynamics can be related to 3-body quantum dynamics in the space
of relative motion with a special class of potentials. As an  important special case the $O(d)$ 
symmetry reduction from $d$ degrees of freedom to one degree is presented. A link between 
two-dimensional (super-integrable) systems and 3-body (super-integrable) systems is revealed. 
As illustration we present number of examples.  We demonstrate that the celebrated 
Calogero-Wolfes 3-body potential has a unique property: two-dimensional quantum dynamics coincides 
with 3-body quantum dynamics on the line at $d=1$; it is governed by the Tremblay-Turbiner-Winternitz 
potential for parameter $k=3$.

\end{abstract}

\maketitle

\newpage

\section{Introduction}

It is well known that one-dimensional quantum dynamics on the half-line can be connected 
to the quantum dynamics in the space of relative motion of 2-body problem with 3 degrees of freedom
and translation-invariant potential, if the potential depends only on the relative distance 
between bodies. It is the consequence of the fact that center-of-mass motion in the 2-body problem 
can be separated out, see e.g. \cite{LL:1977}. The goal of this paper is to answer the question: 
can two-dimensional quantum dynamics be related to dynamics in the relative space of the 
quantum 3-body problem. The answer is affirmative if the 3-body potential depends on two variables 
which allow  parametrization of the space of relative motion.

The two-dimensional Schr\"odinger operator or, equivalently, the Hamiltonian,
\begin{equation}
\label{H2}
  {\cal H}\ =\ -\De^{(2)}\ + \ V(x,y)\ ,\quad
   \De^{(2)}\ =\ \frac{\pa^2}{\pa{{x}}^2 }\ +\ \frac{\pa^2}{\pa{{y}^2}}\ ,\quad {x,y} \in {\bf R}\ ,
\end{equation}
describes a particle with two degrees of freedom subject to the potential $V(x,y)$. In general, 
the system is non-integrable: there exists no non-trivial differential operator
(integral) that commutes with ${\cal H}$. However, for a certain subclass of potentials $V_i$
the Hamiltonian ${\cal H}$ is integrable: there exists a non-trivial differential operator $I$
for which the commutator $[{\cal H}, I]=0$. In many occasions, the integrability occurs when
variables can be separated: this property implies the existence of a first (or second) order
integral in the form of a first (or second) order differential operator.
In particular, if the quantum system is $O(2)$ symmetric, i.e. the potential $V$ is azimuthally
symmetric, variables are separated in polar coordinates and the angular momentum $\hat {\bf L}$
commutes with the Hamiltonian (\ref{H2}). Furthermore, among integrable potentials $V_i$ 
there exists a distinguished class of potentials $V_{si}$, called {\it superintegrable}, 
for which not one but {\it two} integrals $I, {\tilde I}$ occur - the maximum possible 
number of algebraically independent integrals in two dimensions - in addition to the Hamiltonian.
Necessarily, the integrals do not commute, $[I,{\tilde I}]\neq 0$.
This class of potentials has been the subject of numerous studies, for review see \cite{MPW:2013}:
four families of potentials, the so-called Smorodinsky-Winternitz potentials, were introduced
\cite{Fris:1965,Wint:1966}, see also \cite{TTW:2001}, a complete classification has been established
when both integrals are of second order (see \cite{KKM:2005}, \cite{Daskaloyannis} and references
therein). A number of examples is known where in addition to the first-second order integral
there exists a higher order integral, e.g.  \cite{MPW:2013}  
\footnote{There are also examples where both integrals are of order of larger than two, see e.g. 
\cite{PW:2011}}.
Related to this an outstanding property was conjectured \cite{TTW:2001}: any two-dimensional
(any $n$-dimensional, in general) maximally superintegrable quantum system in Euclidean space is 
also exactly-solvable, no single counter-example has been found so far.

The aim of this work is to elaborate a connection between two-dimensional quantum dynamics
and dynamics of relative motion of a 3-body quantum problem. It leads to a relation between
two-dimensional potentials and a special class of translation-invariant 3-body potentials.
In particular, it allows us to employ the above-mentioned description of two-dimensional
(super)integrable potentials $V_{i}(V_{si})(x,y)$ to generate (super)integrable translation-invariant
potentials in the space of relative motion for a quantum 3-body problem.
Most of these translation-invariant potentials are of a unusual type: they can {\it not}
be reduced to the sum of pairwise potential terms.

\section{From a $(d_1+d_2)$-dimensional quantum system to  a
two-dimensional one: symmetry reduction}

We consider a certain $(d_1+d_2)$-dimensional quantum system in ${\bf R}^{d_1} \times {\bf R}^{d_2}$
space with $O(d_1)\times O(d_2)$ symmetry described by the Hamiltonian
\begin{equation}
\label{H_d1+d2}
  {\cal H}_2\ =\ -\De_x^{(d_1)}\ -\ \De_y^{(d_2)}\ + \ V_2(x,y)\ ,
\end{equation}
where
\[
   \De_x^{(d_1)}\ =\ \sum_{i=1}^{d_1} \frac{\pa^2}{\pa{{x}_i} \pa{{x}_i}}\ \equiv\
   \frac{\pa^2}{\pa{{\bf x}} \pa{{\bf x}}} \quad ,\quad
   \De_y^{(d_2)}\ =\ \sum_{i=1}^{d_2} \frac{\pa^2}{\pa{{y}_i} \pa{{y}_i}}\ \equiv\
   \frac{\pa^2}{\pa{{\bf y}} \pa{{\bf y}}} \ ,
\]
are $d_1,d_2$-dimensional Laplacians and $$x={\sqrt{x_1^2+\ldots x_{d_1}^2}}\ ,\
y={\sqrt{y_1^2+\ldots y_{d_2}^2}}$$ are radial distances in $x, y$-spaces, respectively.
It is evident this system has two vectorial integrals - it is characterized by two conserved
angular momenta, $\hat{\bf L}_x=-i\,\bf x \wedge \frac{\pa}{\pa{\bf x}}$ and
$\hat{\bf L}_y=-i\,\bf y \wedge \frac{\pa}{\pa{\bf y}}$ of dimensions $d_1$ and $d_2$,
respectively, here $\wedge$ denotes the exterior product.
Indeed, one can find sets of $(d_1-1)$ commuting 1st and second order symmetries (integrals)
in the enveloping algebra of $so(d_1)$, to which we add the Laplacian $\De_x^{(d_1)}$ to form
$d_1$-dimensional commutative algebra. Similarly, we find $(d_2-1)$ commuting symmetries 
(integrals) for the second Laplacian $\De_y^{(d_2)}$.
Thus, the $(d_1+d_2)$-dimensional Hamiltonian system (\ref{H_d1+d2}) admits  $(d_1+d_2-1)$
commuting independent symmetry operators, including the Hamiltonian.
(For explicit computation of the symmetries corresponding to any separable coordinate system
on the sphere $S_x^{(d_1-1)}$, see \cite{KM1986}.) If the potential in (\ref{H_d1+d2}) 
can be written as $V_2(x,y)=V(x)+V(y)$, an extra integral occurs and 
the Hamiltonian system (\ref{H_d1+d2})
is integrable.

We introduce double spherical coordinates
\[
    {\bf x} = \{x, \Om_x\}\ ,\qquad  {\bf y} = \{y, \Om_y\}\ ,
\]
where $\Om_x, \Om_y$ are Euler angles on spheres $S^{(d_1-1)}, S^{(d_2-1)}$, respectively, 
so that the Hamiltonian (\ref{H_d1+d2}) is reduced to
\begin{equation}
\label{H_d1+d2-r}
  {\cal H}_{2,r}\ =\ -\frac{\pa^2}{\pa x^2} - \frac{d_1-1}{x}\,\frac{\pa}{\pa x} - 
  \frac{\De^{S_x^{(d_1-1)}}}{x^2}\ -\ \frac{\pa^2}{\pa y^2} - \frac{d_2-1}{y}\,\frac{\pa}{\pa y} - 
  \frac{\De^{S_y^{(d_2-1)}}}{y^2}\ + \ V_2(x,y)\ .
\end{equation}
Here $\De^{S_x^{(d_1-1)}},\ \De^{S_y^{(d_2-1)}}$ are Laplacians on spheres $S^{(d_1-1)},\  
S^{(d_2-1)}$, respectively. Since the eigenfunctions of the Laplacian on the sphere are 
multivariable spherical harmonics, each Laplacian decomposes its domain into a direct sum 
of eigenspaces, so each can be replaced by its eigenvalue:
\[
  {\cal H}_{2,radial}(x,y)\ =\ -\frac{\pa^2}{\pa x^2} - \frac{d_1-1}{x}\,\frac{\pa}{\pa x} - 
  \frac{L_x (L_x+d_1-2)}{x^2}\ -\
\]
\begin{equation}
\label{H_d1+d2-radial}
 \frac{\pa^2}{\pa y^2} - \frac{d_2-1}{y}\,\frac{\pa}{\pa y} - \frac{L_y (L_y+d_2-2)}{y^2}\ +\ V_2(x,y)\ ,
\end{equation}
where $L_x, L_y$ index the angular momentum eigenvalues in $x$, $y$ spaces, respectively. 
The configuration space is the quadrant $(x,y) \in {\bf R}_+ \times {\bf R}_+$ in the $E_2$ plane. 
Through the gauge rotation of the Hamiltonian (\ref{H_d1+d2-radial}),
\[
  x^{\frac{d_1-1}{2}} y^{\frac{d_2-1}{2}}\ {\cal H}_{2,radial}\ x^{-\frac{d_1-1}{2}} 
  y^{-\frac{d_2-1}{2}} \equiv H_2\ ,
\]
we arrive at a two-dimensional Hamiltonian,
\begin{equation}
\label{H_2}
    H_2\ =\ -\frac{\pa^2}{\pa x^2}\ -\ \frac{\pa^2}{\pa y^2}\ +\ W_2(x,y)\ ,\ (x,y) \in 
    {\bf R}_+ \times {\bf R}_+
\end{equation}
where the new potential $W_2$ has absorbed the singular terms $\sim 1/x^2$ and $\sim 1/y^2$\
\footnote{
Note that for $d=1,3$ and $L_x=L_y=0$ the singular terms vanish, thus, they do not contribute 
to the potential $W_2=V_2$.}. The above-described procedure of $O(d_1) (O(d_2))$ symmetry 
reduction is illustrated by the following concrete examples.

\noindent
{\bf (I).\it\ Two hydrogen atoms.}

In the $1/Z$-expansion for the Helium-like sequence, see e.g. \cite{TLO:2019} and references 
therein, the leading term is described by two three-dimensional hydrogen atoms,
\begin{equation}
\label{H_3+3}
  {\cal H}_2\ =\ -\De_{r_1}^{(3)}\ -\ \De_{r_2}^{(3)}\ + \ V_2({r_1},{r_2})\ ,
      \quad V_2\ =\ -\frac{1}{r_1}-\frac{1}{r_2}\ ,
\end{equation}
cf. (\ref{H_d1+d2}), where
\[
  r_1={\sqrt{x_1^2+x_2^2+ x_{3}^2}}\ ,\  r_2={\sqrt{y_1^2+y_2^2+ y_{3}^2}}\ ,
\]
are radial distances in $r_1, r_2$-spaces, respectively. It is characterized by $O(3)\times O(3)$ 
symmetry and by four vectorial integrals: two 3D angular momenta $\bf L_1 (\bf L_2)$ and 
two 3D Runge-Lenz vectors $\bf A_1 (\bf A_2)$ with constraints $(\bf L_1,\bf A_1)=(\bf L_2,\bf A_2)=0$. 
There is an evident extra integral due to separation of the $r_1, r_2$ variables.
The total number of integrals is equal to eleven.  Needless to say, one can find six integrals 
forming a commutative algebra. This system is maximally-superintegrable.

Making the symmetry reduction at $L_1=L_2=0$ we arrive at the two-dimensional Hamiltonian,
\begin{equation}
\label{H_2-I}
    H_2\ =\ -\frac{\pa^2}{{\pa {r_1}}^2}\ -\ \frac{\pa^2}{{\pa {r_2}}^2}\ +\ V_2(r_1,r_2)
    \equiv\ H(r_1)+H(r_2)\ ,\quad (r_1,r_2) \in {\bf R}_+ \times {\bf R}_+ \ ,
\end{equation}
which defines integrable system.

\bigskip

\noindent
{\bf (II).\it\ Harmonic oscillator.}

Let us consider $(d_1+d_2)$-dimensional anisotropic harmonic oscillator with singular terms in 
${\bf R}^{d_1} \times {\bf R}^{d_2}$ space with $O(d_1)\times O(d_2)$ symmetry. It is described 
by the Hamiltonian
\begin{equation}
\label{H_d1+d2-II}
  {\cal H}_2\ =\ -\De_x^{(d_1)}\ -\ \De_y^{(d_2)}\ + \ a\, \om^2\, x^2\ +\
  b\,\om^2\, y^2\ +\ \frac{A}{x^2}\ +\ \frac{B}{y^2}\ ,
\end{equation}
where $a, b, \om, A, B$ are parameters, $x,y$ are radial distances in $d_1, d_2$-dimensional spaces.
Making symmetry reduction we arrive at the caged harmonic oscillator in two dimensions \cite{EV:2008}
\begin{equation}
\label{H_2-II}
    H_2\ =\ -\frac{\pa^2}{\pa x^2}\ -\ \frac{\pa^2}{\pa y^2}\ +\ a\,\om^2\, x^2\ +\ 
    b\,\om^2\, y^2\ +\ \frac{\hat A}{x^2}\ +\ \frac{\hat B}{y^2}\ \equiv\ H(x)+H(y)
    \ ,\quad (x,y) \in {\bf R}_+ \times {\bf R}_+\ .
\end{equation}
Due to separation of variables the Hamiltonian $H_2$ defines the integrable system,
$[H_2, H(x)]=0$. If $a,b, \om$ are positive and $A,B > -\frac{1}{8}$, this system is also 
exactly-solvable with spectra linear in quantum numbers. Its hidden algebra is $sl_2 \oplus sl_2$.

Interestingly, there are two particular cases of (\ref{H_2-II}) discovered in \cite{Fris:1965,Wint:1966}
\[
  a\ =\ b\ =\ 1\ ,
\]
\[
   a\ =\ 4\ ,\ b\ =\ 1\ ,\ {\hat A}\ =\ 0\ ,
\]
where this system becomes superintegrable: there exists an additional second order integral. 
In both cases the hidden algebra becomes $sl_2$. In \cite{JH:1940,EV:2008} it was found that 
if $a=N^2, b=1$, so that the potential takes the form
\begin{equation}
\label{9a}
 \tilde W_N \ =\ \om^2\,(N^2\,x^2 + y^2)\ +\ \frac{\hat A}{x^2}\ +\ \frac{\hat B}{y^2} \ ,
\end{equation}
the system is also superintegrable with a second integral of the $N$th order.

\vskip 0.6cm

\noindent
{\bf (III). Tremblay-Turbiner-Winternitz (TTW) and Post-Winternitz (PW) potentials.}

In 2009 there was discovered a few parametric potential \cite{TTW:2009}, called the TTW potential,
written easily in polar coordinates $(\rho, \theta)$
\begin{eqnarray}
  V_{\rm TTW}^{(k)} \ = \ \om^2\, \rho^2 \ + \ \frac{1}{\rho^2} \left[\frac{\al}{\cos^2 (k\, \theta)} +
  \frac{\beta}{\sin^2 (k \,\theta)}\right]\ ,
\label{TTW-pot}
\end{eqnarray}
that admits two non-commutative integrals,
one of the second order and another one of the $(N=k)$th order in the case of integer $k$.
Here $\al, \beta, \om$
are free parameters. In Cartesian coordinates $(x,y)$ the singular terms of the potential
are represented by a rational function.

If $k$ is rational, $k=m/n$ with $m$ and $n$ integers (with no common divisors)
the additional $N$-th order integral \cite{QuesneTTW:2010,KalninsKM:2010}
\begin{eqnarray}
      N \ =\  2 \,(m \,+\, n\, -\, 1)\ ,
\label{TTW-N}
\end{eqnarray}
occurs. For $m=n=1$, the TTW system possesses two 2nd order integrals (as the consequence
of multiseparability of variables in Cartesian and polar coordinates), it coincides with the
caged oscillator (\ref{H_2-II}) at $a=b=1$.

Another higher order superintegrable system separating in polar coordinates has
potential,
\begin{eqnarray}
    V_{\rm PW}^{(k)}\ = \ -\frac{a}{\rho}\ + \
    \frac{1}{\rho^2}\left[\frac{\mu}{\cos^2(\frac{k}{2} \theta)}\ +
    \ \frac{\nu}{\sin^2(\frac{k}{2} \theta)} \right]\ ,
\label{PW}
\end{eqnarray}
see \cite{PW:2010}, with $k=m/n$ being rational. This potential called the PW potential,
it is related to TTW by coupling constant metamorphosis \cite{MillerPost}.


\section{From a $3$-body system with $d$-degrees of freedom \\to  a $2$-dimensional one}

Now let us consider the $3$-body quantum system of $d$ degrees of freedom with masses
$m_1,m_2,m_3$ and translation-invariant potential $V_r$. Its kinetic energy is of the form,
\begin{equation}
\label{Tflat}
   {\cal T}\ =\ -\sum_{i=1}^3 \frac{1}{2m_i}\De_i^{(d)}\ ,
\end{equation}
with coordinate vector of $i$th particle ${\bf r}_i \equiv {\bf r}^{(d)}_i=(x_{i,1}\,,
\cdots \,,x_{i,d})$. Here, $\De_i^{(d)}$ is the $d$-dimensional Laplacian,
\[
     \De_i^{(d)}\ =\ \frac{\pa^2}{\pa{{\bf r}_i} \pa{{\bf r}_i}}\ ,
\]
associated with the $i$th particle. Evidently, its $3d$-dimensional configuration space is
${\bf R}^{d} \times {\bf R}^{d}\times {\bf R}^{d}$ space and (\ref{Tflat}) is
$O(d)\times O(d)\times O(d)$ symmetric. This system is characterized by three
conserved angular momenta, $\hat{\bf L}_{1,2,3}$, respectively.

We introduce the center-of-mass, $d$-dimensional vectorial coordinate,
\begin{equation}
\label{CMS}
    {\bf R}_{0} \ =\ \frac{1}{\sqrt{M}}\,\sum_{k=1}^{3} m_k {\bf r}_{_k}\ ,
    \ M=m_1+m_2+m_3\ ,
\end{equation}
where $M$ is the total mass, and two, $d$-dimensional, vectorial Jacobi coordinates
\begin{equation}
\label{Jacobi}
    {\bf r}^{(J)}_{1} \ = \ \sqrt{\frac{m_1 m_2}{m_1+m_2}}({\bf r}_{2}-{\bf r}_{1})\ ,\
        {\bf r}^{(J)}_{2} \ = \ \sqrt{\frac{m_3 (m_1+m_2)}{M}}\left({\bf r}_{3}-
        \frac{m_1}{m_1+m_2}\,{\bf r}_1-\frac{m_2}{m_1+m_2}\,{\bf r}_2 \right)\  ,
\end{equation}
see e.g. \cite{Delves:1960} and also \cite{TME3-d} for discussion, which are translation-invariant.
In Jacobi coordinates the Laplacian (\ref{Tflat}) becomes diagonal,
\begin{equation}
\label{Tflat-diag}
   {\cal T}\ =\ -\frac{\pa^2}{\pa{{\bf R}_0} \pa{{\bf R}_0}}\ - \
       \frac{\pa^2}{\pa{{\bf r}_1^{(J)}} \pa{{\bf r}_1^{(J)}}}\ -\
       \frac{\pa^2}{\pa{{\bf r}_2^{(J)}}
       \pa{{\bf r}_2^{(J)}}} \equiv \ {\cal T}_0\ + {\cal T}_r\ ,
\end{equation}
where ${\cal T}_0=-\De_{{\bf R}_0}$ is the kinetic energy of the center-of-mass motion,
while ${\cal T}_r$ is the kinetic energy of relative motion.

Due to translation-invariance of 3-body potential $V_r$, there is no dependence on
the center-of-mass coordinate ${\bf R}_0$. We restrict our consideration to
the potentials which depend on Jacobi distances $$r_i^{(J)}=|{\bf r}_i^{(J)}|, i=1,2\ ,$$
{\it only}. Separating out the ${\bf R}_0$-coordinate, we arrive at the Hamiltonian
of the relative motion,
\begin{equation}
\label{Hrel}
  {\cal H}_r\ =\ -\De_1^{(d)}\ -\ \De_2^{(d)}\ + \ V_r(r_1^{(J)},r_2^{(J)})\ ,
\end{equation}
cf.(\ref{H_d1+d2}). Here
$$\De_i^{(d)}=\frac{\pa^2}{\pa{{\bf r}_i^{(J)}} \pa{{\bf r}_i^{(J)}}}\ ,\ i=1,2\ ,$$
are the Laplacians.

The $(2d)$-dimensional system described by (\ref{Hrel}) is $O(d)\times O(d)$ symmetric.
Two Jacobi distances $r_1^{(J)}, r_2^{(J)}$ play a role of radii. The system
is characterized by two conserved angular momenta,
$\hat{\bf L}_1=-i\,{\bf r}^{(J)}_{1} \wedge \frac{\pa}{\pa{{\bf r}^{(J)}_{1}}}$
and $\hat{\bf L}_2=-i\,{\bf r}^{(J)}_{2} \wedge \frac{\pa}{\pa{{\bf r}^{(J)}_{2}}}$.

Now one can make the $O(d)$ symmetry reduction of Section II for the Hamiltonian (\ref{Hrel}) 
to the two-dimensional case in a similar manner to that which was done for the Hamiltonian 
(\ref{H_d1+d2}).
In terms of double spherical coordinates
\[
    {\bf r}_1^{(J)}\ =\ \{r_1^{(J)}, \Om_1\}\ ,\  {\bf r}_2^{(J)}\ =\ \{r_2^{(J)}, \Om_2\}\ ,
\]
where $\Om_1, \Om_2$ are Euler angles on spheres $S^{(d-1)}_{1,2}$, the Hamiltonian (\ref{Hrel})
is reduced to the double radial-spherical operator
\begin{equation}
\label{H_d-r12}
  {\cal H}_{r}\ =\ -\frac{\pa^2}{(\pa r_1^{(J)})^2} - \frac{d-1}{r_1^{(J)}}\,\frac{\pa}{\pa r_1^{(J)}}
  - \frac{\De^{S_1^{(d-1)}}}{(r_1^{(J)})^2}\ -\ \frac{\pa^2}{(\pa r_2^{(J)})^2} -
  \frac{d-1}{r_2^{(J)}}\,\frac{\pa}{\pa r_2^{(J)}} - \frac{\De^{S_2^{(d-1)}}}{(r_2^{(J)})^2}\ +
  \ V_r(r_1^{(J)},r_2^{(J)})\ .
\end{equation}
cf.(3).
Here $\De^{S_{1,2}^{(d-1)}}$ are Laplacians on spheres $S_{1,2}^{(d-1)}$, respectively.
Since the eigenfunctions of the Laplacian on the sphere are spherical harmonics,
each Laplacian can be replaced by an  eigenvalue,
\[
  {\cal H}_{r,radial}(x,y)\ =\ -\frac{\pa^2}{(\pa r_1^{(J)})^2}\ -\
  \frac{d-1}{r_1^{(J)}}\,\frac{\pa}{\pa r_1^{(J)}}\ -\ \frac{L_1 (L_1+d-2)}{(r_1^{(J)})^2}\ -\
\]
\begin{equation}
\label{H_d-r12-radial}
 \frac{\pa^2}{(\pa r_2^{(J)})^2}\ -\ \frac{d-1}{r_2^{(J)}}\,\frac{\pa}{\pa r_2^{(J)}}\ -\
 \frac{L_2 (L_2+d-2)}{(r_2^{(J)})^2}\ +\ V_r(r_1^{(J)},r_2^{(J)})\ ,
\end{equation}
where $L_1, L_2$ index the angular momenta in $1,2$ spaces, respectively.
The configuration space is the quadrant ${\bf R}_+(1) \times {\bf R}_+(2)$ in the $E_2$ plane.
Through the gauge rotation of the Hamiltonian (\ref{H_d-r12-radial}),
\[
  ({r_1^{(J)}}{r_2^{(J)}})^{\frac{d-1}{2}} \ {\cal H}_{r,radial}\
  ({r_1^{(J)}}{r_2^{(J)}})^{-\frac{d-1}{2}} \equiv H_r\ ,
\]
we arrive at the two-dimensional Hamiltonian,
\begin{equation}
\label{H_r}
    H_r\ =\ -\frac{\pa^2}{(\pa r_1^{(J)})^2}\ -\ \frac{\pa^2}{(\pa r_2^{(J)})^2}\ +\
    W_r(r_1^{(J)},r_2^{(J)})\ ,
\end{equation}
cf.(\ref{H_2}), where the new potential $W_r$ has absorbed the singular terms
$\sim 1/(r_1^{(J)})^2$ and $\sim 1/(r_2^{(J)})^2$. It is worth noting that for $d=1,3$ and $L_1=L_2=0$
the singular terms are not added to new potential and
\[
    V_r\ =\ W_r\ .
\]

It can be immediately seen that by identifying $(r_1^{(J)},r_2^{(J)})$ with $(x,y)$,
\begin{equation}
\label{trans}
       x \rar r_1^{(J)}\ ,\ y \rar r_2^{(J)}\ .
\end{equation}
the Hamiltonian $H_r$ (\ref{H_r}) of the relative motion becomes the two-dimensional Hamiltonian
$H_2$ (\ref{H_2}). This identification allows us to transform the theory of (super)integrable 
two-dimensional systems in the $(x,y)$-plane to the theory of (super)integrable systems in the 
space of relative motion of the 3-body problem parameterized by Jacobi distances 
$(r_1^{(J)}),(r_2^{(J)})$,

In particular, the integrable potential (\ref{H_3+3}) is transformed to a 3-body two-dimensional 
integrable potential in the plane of relative motion
\[
  V_2\ =\ -\frac{1}{r_1}-\frac{1}{r_2}\ \rar\ W_r\ =\ -\frac{1}{r_1^{(J)}}-\frac{1}{r_2^{(J)}}\ .
\]
It is easy to check that the two-dimensional anisotropic harmonic oscillator (\ref{H_2-II}) at
$\hat{A}=\hat{B}=0$ becomes the 3-body Jacobi oscillator \cite{TWE:2020},
\[
  W_2 = a\, \om^2\, x^2\ +\ b\,\om^2\, y^2 \ \rar\
  W_r = a\, \om^2\,(r_1^{(J)})^2 +\  b\,\om^2\,(r_2^{(J)})^2 \ .
\]
In general, the superintegrable caged oscillator (\ref{9a}) becomes the 3-body superintegrable
\textit{caged} Jacobi oscillator
\[
   W_r= a\, \om^2\,(r_1^{(J)})^2 +\  b\,\om^2\,(r_2^{(J)})^2\ +\
   \frac{A}{(r_1^{(J)})^2}\ +\ \frac{B}{(r_2^{(J)})^2}\ .
\]
Similarly, the 3-body analogue of the TTW potential (\ref{TTW-pot}) is of the form
\begin{equation}
\label{TTW1}
    W_r^{(k)} (\rho_r, \tha_r; \om, \al, \beta)\ =\ \om^2 {\rho_r}^2 +
    \frac{\al k^2}{{\rho_r}^2 \cos^2 {k \tha_r}} + \frac{\beta k^2}{{\rho_r}^2 \sin^2 {k \tha_r}}\ ,
\end{equation}
being written in Jacobi polar coordinates $(\rho_r\equiv {\rho}_r^{(J)}, \tha_r \equiv 
{\tha}_r^{(J)})$ instead of Jacobi distances
$(r_1^{(J)},r_2^{(J)})$, where $\rho^{(J)}_r=\sqrt {(r_1^{(J)})^2+(r_2^{(J)})^2}$ and
$\tha^{(J)}_r$ is polar angle. In these coordinates the 3-body TTW Hamiltonian admits separation
of variables in the plane of relative motion. Here $\al, \beta > - \frac{1}{4 k^2}$, $\om$ and
$k \neq 0$ are parameters. Correspondingly, the Hamiltonian,
\[
    {\cal H}_r\ =\ {\cal T}\ +\ W_r^{(k)} \ ,\quad k=m/n\ ,
\]
see (\ref{Tflat}), presents one of the most striking examples of 3-body superintegrable
systems with $2(m \,+\, n\, -\, 1)${\it th} order extra integral of motions, see
(\ref{TTW-N}).

At $d=1$ the general 3-body potential becomes two-dimensional: it depends on two independent
variables {\it only}, the triangle of interaction with sides $r_{12},r_{13},r_{23}$ degenerates
to the interval with marked point. If body positions are ordered like $r_1 \leq r_2 \leq r_3$,
the relative distances are constrained,
\[
    r_{13}\ =\ r_{12}+r_{23}\ ,
\]
and the Jacobi distances (\ref{Jacobi}) at $m_1=m_2=m_3=2$ are expressed as,
\[
   r_1^{(J)}=r_{12}\ ,\ r_2^{(J)}={\sqrt\frac{4}{3}}(r_{13}+r_{23})\ .
\]
In this case one can relate the 3-body relative motion dynamics to the two-dimensional one:
\begin{equation}
\label{trans-back}
  (r_1^{(J)},r_2^{(J)})\ \rar\ (x,y)\ .
\end{equation}

There are two remarkable examples of superintegrable and exact-solvable 3-body problem at $d=1$:
the Calogero model \cite{Calogero:1969} with pairwise interaction with potential
\begin{equation}
\label{Calogero}
  W^{(c)}_2 = \om^2 (r_{12}^2+r_{23}^2 + r_{13}^2) +
  A \left(\frac{1}{r_{12}^2} + \frac{1}{r_{23}^2}+ \frac{1}{r_{13}^2} \right)\ ,
\end{equation}
and, which is more general, the Wolfes model \cite{Wolfes:1974} with 2 and 3 body interactions
\begin{equation}
\label{Wolfes}
  W^{(w)}_2 = W^{(c)}_2\ +\ B\,\left(\frac{1}{(r_{12}+r_{13})^2} + \frac{1}{(r_{12}+r_{23})^2}+
  \frac{1}{(r_{13}+r_{23})^2} \right)\ .
\end{equation}
It can be easily shown that in Jacobi distances (parametrized by polar coordinates
$({\rho}_r^{(J)}, {\tha}_r^{(J)})$, see above) the Wolfes potential becomes the TTW potential at $k=3$,
\begin{equation}
\label{TTW-3body}
    W^{(w)}_2 = W_r^{(3)} (\rho_r, \tha_r; \om, \al, \beta)\ =\ \om^2 {\rho_r}^2 +
    \frac{9 \al }{{\rho_r}^2 \cos^2 {3 \tha_r}} + \frac{9\beta}{{\rho_r}^2 \sin^2 {3 \tha_r}}\ ,
\end{equation}
see (\ref{TTW1}). Making identification (\ref{trans-back}) we arrive at the two-dimensional potential
\begin{equation}
\label{TTW-2D}
    V^{(w)}_2 = V_{TTW}^{(3)} (\rho, \tha; \om, \al, \beta)\ =\ \om^2 {\rho}^2 +
    \frac{9 \al }{{\rho}^2 \cos^2 {3 \tha}} + \frac{9\beta}{{\rho}^2 \sin^2 {3 \tha}}\ ,
\end{equation}
cf.(\ref{TTW-pot}) written in polar coordinates $(\rho,\tha)$.

%
%

\section{Conclusions and discussion}

We have established a quantum mechanical construction linking planar dynamics to one related
to 3-body relative motion. It is evident that this construction can be extended even further
by making a connection of the $n$-dimensional Schr\"odinger equation to an $(n+1)$-body quantum
system by identifying the Cartesian coordinates with Jacobi distances. This allows us to
connect integrability/solvability properties of quantum dynamics in Cartesian coordinates
with the quantum dynamics of relative motion in the many-body problem. It has an obvious
classical mechanical analog for all the potentials that we have presented. In particular,
in \cite{KalninsKressMiller} it is shown how to construct additional classical
superintegrable 2D Euclidean systems with higher order integrals, following the TTW method
\cite{TTW:2010}. It has not yet been established  which of these new systems is also
quantum superintegrable. This will be done elsewhere.

\section*{Acknowledgments}

A.V.T. is thankful to University of Minnesota, USA for kind hospitality extended to him
where this work was initiated and where it was mostly completed.
W.M. was partially supported by a grant from the Simons Foundation (\# 412351
to Willard Miller, Jr.). M.A.E.R. is grateful to ICN-UNAM (Mexico) for kind hospitality
extended to him where essential part of this work was done during his numerous visits.
This research is partially supported by DGAPA IN113819 and
CONACyT A1-S-17364 grants (Mexico).



\begin{thebibliography}{99}

\bibitem{LL:1977}
     L.D.~Landau and E.M.~Lifshitz,\\
     {\it Quantum Mechanics, Non-relativistic Theory} {\rm (Course of Theoretical Physics, vol.3)},\\
     {3rd edn (Oxford:Pergamon Press)}, 1977

\bibitem{MPW:2013}
        W.~Miller,~Jr., S.~Post and P.~Winternitz,\\
        {\it J. Phys. \bf A46} (2013) 423001

\bibitem{Fris:1965}
        J.~Fri$\breve{s}$, V.~Mandrosov, Ya.A.~Smorodinsky, M.~Uhlir and P.~Winternitz,\\
        {\it Phys.Lett. \bf 16}, 354 (1965)

\bibitem{Wint:1966}
        P.~Winternitz, Ya.A.~Smorodinsky, M.~Uhlir and J.~Fri$\breve{s}$,\\
        {\it Yad. Fiz. \bf 4}, 625 (1966);\\
        \emph{Sov.Journ.Nucl.Phys. \textbf{4}}, 444 (1967)\ (English Translation)

\bibitem{TTW:2001}
        P.~Tempesta, A.V.~Turbiner and P.~Winternitz,\\
        \textit{Exact Solvability of Superintegrable Systems},\\
        {\it J. Math. Phys. \bf 42}, 4248-4258 (2001)

\bibitem{KKM:2005}
        E.G.~Kalnins,  J.M.~Kress and  W.~Miller Jr.,\\
        {\it J. Math. Phys. \bf 46} (2005) 053510

\bibitem{Daskaloyannis}
         C.~Daskaloyannis and  Y.~Tanoudis,\\
        {\it J. Math. Phys. \bf 48} (2007) 072108

\bibitem{PW:2011}
         S.~Post and  P.~Winternitz,\\
         {\it J. Phys. \bf A44}  (2011) 162001

\bibitem{TTW:2009}
        F.~Tremblay, A.V.~Turbiner and P.~Winternitz,\\
        {\it J. Phys. \bf A42} (2009) 242001

\bibitem{PW:2010}
          S.~Post and P.~Winternitz,\\
          {\it J. Phys. \bf A43} (2010) 222001

\bibitem{KM1986}
         E.G.~Kalnins and W.~Miller Jr., \\
         {\it J. Math. Phys. \bf 27} (1986) 1721-1736

\bibitem{TLO:2019}
         A.V.~Turbiner, J.C.~Lopez Vieyra and H.~Olivares Pilon\\
         {\it Ann. Phys. \bf 409} (2019) 167908 

\bibitem{JH:1940}
        J.M.~Jauch and E.L.~Hill,\\
        {\it Phys. Rev. \bf 57}, 641 (1940)


\bibitem{EV:2008}
        N.W.~Evans and P.E.~Verrier,\\
        {\it J. Math. Phys. \bf 49}, 092902 (2008)


\bibitem{QuesneTTW:2010}
          C.~Quesne,\\
         {\it J. Phys. \bf A43} 082001 (2010)

\bibitem{KalninsKM:2010}
          E.G.~Kalnins,  J.M.~Kress and  W.~Miller Jr.,\\
         {\it J. Phys. \bf A43} 265205 (2010)

\bibitem{MillerPost}
         E.G.~Kalnins, S.~Post and  W.~Miller Jr.,\\
         {\it J. Phys. \bf A43}  035202 (2010)


\bibitem{Delves:1960}
        L.M.~Delves,\\
        {\it Nucl. Phys. \bf 20}, 275-308 (1960)

\bibitem{TME3-d}
         A.~Turbiner, W.~Miller~Jr. and M.~A.~Escobar-Ruiz,\\
         {\it J. Math. Phys. \bf 59} (2018) 022108 

\bibitem{TWE:2020}
         A.~Turbiner, M.A.~Escobar-Ruiz and W.~Miller~Jr.\\
         {\it J. Phys. \bf A53} (2020) 055302 

\bibitem{Calogero:1969}
           F.~Calogero,\\
           {\em J.~Math. Phys. \bf 10},  2191--2196 (1969)

\bibitem{Wolfes:1974}
           J.~Wolfes,\\
           {\em J.~Math. Phys. \bf 15}, 1420-1424  (1974)

\bibitem{KalninsKressMiller}
          E.G.~Kalnins,  J.M.~Kress and W.~Miller Jr.,\\
         {\it SIGMA} {\bf 6} (2010) 066

\bibitem{TTW:2010}
         F.~Tremblay, A.V.~Turbiner and P.~Winternitz,\\
        {\it J. Phys. \bf A43} (2010) 015202 

\end{thebibliography}
\end{document}